\documentclass[preprint,review,12pt]{elsarticle}
\journal{Physica C}

\begin{document}


\title{Effect of the Fermi surface reconstruction on the self-energy of the
copper-oxide superconductors
}


\author[1]{B. Bellafi}
\author[2]{M. Azzouz}
\ead{mazzouz@laurentian.ca}
\author[1]{S. Charfi-Kaddour}
\address[1]{
Laboratoire de Physique de la Mati\`ere Condens\'ee, D\'epartement de Physique, 
Facult\'e des Sciences de Tunis, Campus Universitaire 1060 Tunis, Tunisia}
\address[2]{Department of Physics, Laurentian University,
Ramsey Lake Road, Sudbury, Ontario, Canada P3E 2C6}
%



\begin{abstract}
We calculated the self-energy corrections beyond the 
mean-field solution of the rotating antiferromagnetism 
theory using the functional integral approach.  
The frequency dependence of the scattering rate 
${1}/{\tau}$ 
is evaluated for different temperatures and doping levels, 
and is compared with other 
approaches and with experiment. 
The general trends we found are fairly consistent with the extended 
Drude analysis of the optical conductivity, and with 
the nearly antiferromagnetic Fermi liquid as far as the ${\bf k}$-anisotropy 
is concerned and some aspects of the Marginal-Fermi liquid behavior. 
The present approach
provides the justification from the microscopic point of view for the 
phenomenology of the marginal Fermi liquid
ansatz, which was used in the calculation of several physical
properties of the high-$T_C$ cuprates
within the rotating antiferromagnetism theory. In addition, the expression of
self-energy we calculated takes into account
the two hot issues of the 
high-$T_C$ cuprate
superconductors, namely the Fermi surface reconstruction 
and the hidden symmetry, 
which we believe are related to the pseudogap.
\end{abstract}
\maketitle

\section{Introduction}

The origin of the pseudogap  (PG) \cite{timusk1999} 
behavior of the high-$T_C$ cuprate superconductors (HTSC) 
remains an open issue even though more than quarter a century 
has passed after the discovery of superconductivity in 
these materials \cite{bednorz1986}.
The PG phase turned out to be more challenging and subtle 
than the superconducting
phase itself. Indeed, the PG has been measured as 
a depression in the density of states at 
the Fermi energy
below the doping dependent PG temperature $T^*$, but
no broken symmetry has so far been observed beyond any doubt \cite{he2011}.
A number of theoretical models have been proposed 
in order to explain this PG phenomenon, with some based 
on the preformed-pairs scenario and others 
based on competing orders \cite{chakravarty2001}.
The rotating antiferromagnetism theory (RAFT), which
belongs in the latter,
is characterized by two competing orders; namely the d-wave 
superconductivity and the rotating antiferromagnetic (RAF) order.
The RAF order parameter has a finite magnitude below 
a temperature, which was identified with $T^*$, and a phase that varies with 
time \cite{azzouz2003,azzouz2003p,azzouz2004,azzouz2012p,azzouz2013,azzouz2013p}.
RAFT yield results in good agreement with 
several experimental data of the HTSCs.
Resistivity \cite{azzouz2005}, optical 
conductivity \cite{azzouz2012}, Raman \cite{azzouz2010},
and ARPES \cite{azzouz2010p,azzouz2013} have been analyzed within RAFT
assuming the phenomenological marginal-Fermi liquid 
(MFL) self-energy \cite{varma1989}. Until before the 
completion of this work, the justification for using
this assumption was missing.
The results of this work show that going beyond the mean-field 
solution of RAFT a self energy that is 
consistent with a MFL is derived.
More importantly, in the limit of the tight-binding bare electrons
our self-energy satisfies the same equation as in 
the second-order Born approximation, which was used in 
the nearly antiferromagnetic Fermi liquid (NAFL) 
theory \cite{stojkovic1997}.
Moreover, we generalize this approximation  
into a gapped second-order Born approximation that takes 
into consideration the PG.
Interestingly,
we can qualify the RAF
state as a state that is nearly antiferromagnetic
because the RAF state
has the same (free) energy as a true 
ordered antiferromagnetic state but is a disordered state because of
the time dependence of the phase of the RAF
order parameter.

Below the PG temperature in the underdoped regime,
we find that the relaxation rate displays a linear behavior
at large frequencies consistent with a marginal Fermi liquid, 
but it displays strong deviation from linearity at low
frequencies, which is characterized by a hump due to the PG.
In the overdoped regime at any temperature or above the PG
temperature in the underdoped regime, the relaxation rate 
shows a mixture of Fermi liquid (FL) and MFL behaviors.
We argue that this evolution with doping is related to 
the Fermi surface (FS) reconstruction \cite{azzouz2013}.

This work is organized as follows. In Sec. \ref{sec2}
we calculate the Gaussian corrections to the
mean-field solution of RAFT using a Hubbard-Stratanovich
identity that decouples the quartic
term of the Hubbard model in the channel 
of RAF order. This yields 
the propagator of the Gaussian fluctuations.
Self-energy is calculated in Sec. \ref{sec3} using this propagator, and 
a gapped second-order Born approximation
is derived for self-energy in the presence of the PG.
Some numerical results are presented in Sec. \ref{sec4}, and
conclusions are drawn in Sec. \ref{sec5}.

\section{Method}
\label{sec2}

RAFT has been developed using the
extended Hubbard model, with a repulsive on-site 
Coulomb interaction and a nearest-neighbor attractive 
interaction that simulates d-wave pairing. 
Here, we focus on the normal (non superconducting)
state, where the Hamiltonian on a two-dimensional lattice reads as
\begin{eqnarray}
\label{hamiltonian}
H &=& H_0 + H_I\cr
&=&
-t\sum_{\langle i,j\rangle,\sigma} c^{\dag}_{i,\sigma} c_{j,\sigma} 
-t^{\prime}\sum_{\langle\langle i,j\rangle\rangle,\sigma} 
c^{\dag}_{i,\sigma} c_{j,\sigma} \cr
&& - \mu \sum_{i,\sigma} c^{\dag}_{i,\sigma} c_{j,\sigma} 
+ U\sum_{i}n_{i,\uparrow}n_{i,\downarrow}.
\end{eqnarray}
In (\ref{hamiltonian}), 
$H_0$ stands for the kinetic and chemical potential energies, and 
$H_I=U\sum_i n_{i\uparrow}n_{i\downarrow}$
is the sum of all on-site Coulomb energies.
$t$ and $t^{\prime}$ designate the electron's hopping 
energies between the nearest-neighbor $(\langle i,j\rangle)$ 
and next-nearest-neighbor 
$(\langle\langle i,j\rangle\rangle)$ sites respectively, 
$\mu$ is the chemical potential, $c^{\dag}_{i,\sigma}$ 
($c_{j,\sigma}$) creates (annihilates) an electron with 
spin $\sigma$ at site $i$, 
and $n_{i,\sigma} = c^{\dag}_{i,\sigma} c_{i,\sigma}$ 
is the number operator. 

The partition function can be written as \cite{negele}
\begin{eqnarray}
 Z=\int \prod_{i,\sigma}dc_{i,\sigma}^* dc_{i,\sigma}\,
e^{-\int_{0}^{\beta}\,d\tau[\Sigma_{i,\sigma}
c_{i,\sigma}^*\frac{\partial}{\partial\tau}c_{i,\sigma}\,+H_0+H_{I}]},
\end{eqnarray}
where $c$ and $c^*$ are from now on anticommuting Grassmann variables.
For the RAF order, we decouple 
the interacting $U$-term of (\ref{hamiltonian}) 
using a Hubbard-Stratanovich transformation by considering 
the RAF order parameter 
$Q=\langle c_{i,\sigma}c^\dag_{i,\sigma}\rangle$, 
which has been used to model the PG
behavior \cite{azzouz2003, azzouz2003p}. This gives
\begin{eqnarray}
\label{HS}
e^{-\int d\tau H_{I}}&=&\int \prod_i db_{i} 
\exp\{ \int d\tau [-\sum_{i}b_i^{*}U^{-1}b_{i} \cr
&&+\sum_{i}c_{i\downarrow}c_{i\uparrow}^{*}b_{i}
+\sum_{i}c_{i\uparrow}c_{i\downarrow}^{*}b_{i}^{*}]\},
\end{eqnarray}
where $b_i$ is a Hubbard-Stratanovich complex field.
In order to recover the RAF state 
at the mean-field level in the present treatment we write the field $b_i$
as
\begin{equation}
b_i=\mid b_i \mid e^{i[\pi(x_i+y_i)+\phi(t)]}.
\end{equation}
The phase term $e^{i\pi(x_i+y_i)}=(-1)^{x_i+y_i}$ guarantees that the rotating
order parameter is staggered due to 
the antiferromagnetic correlations, 
and the time-dependent phase $\phi(t)$ 
insures that the staggered magnetization 
rotates \cite{azzouz2013p,azzouz2012,azzouz2003p,azzouz2004}.
Using the Grassmann variables and the transformation (\ref{HS}), 
the partition function takes on the form
\begin{equation}
Z=\int\prod_{i,\sigma}dc_{i,\sigma}^* dc_{i,\sigma}db_i \exp(-S_{eff}),
\end{equation}
with 
\begin{eqnarray}
S_{eff}&&=\int_{0}^{\beta}\,d\tau[\sum_{i\sigma}
\sum_{\alpha=A,B}{c_{i \sigma}^{\alpha^*}
\frac{\partial}{\partial\tau}c_{i,\sigma}^{\alpha}}+H_0^{\alpha}\cr
&&+\sum_{i;\alpha=A,B}(c_{i\uparrow}^{\alpha^*}
c_{i\downarrow}^{\alpha}b_{i}+
c_{i\downarrow}^{\alpha^*}c_{i\uparrow}^{\alpha}b^*_{i})
+\sum_{i}\frac{\mid b_i\mid^2}{U}].
\end{eqnarray}
Here 
$\beta=\frac{1}{k_BT}$ is inverse temperature, and $A$ and $B$ 
designate the two sublattices of the bipartite lattice.
The upper index $\alpha$ in $H_0^\alpha$ means that the single 
particle part of the Hamiltonian has now to be written 
using the two sublattices, $A$ and $B$.
The mean-field solution, where $b_i\equiv b_0$ is time and 
space independent, allows us to recover the RAFT's 
mean field equation for the parameter 
$Q=|\langle c_{i,\uparrow}c_{i,\downarrow}^\dag\rangle|$, Ref. \cite{azzouz2003}:
\begin{equation}
1=\frac{U}{2N} \sum_{\bf k} \frac{n_F[E_-({\bf k})]-n_F[E_+({\bf k})]}{E_{q}({\bf k})},
\label{mean-field equation}
\end{equation}
where $n_F(E) = \frac{1}{1 + e^{\beta E}}$, $E_{q}({\bf k}) 
= \sqrt{\epsilon_1({\bf k})^2 +b_0^2}$,
and $N$ is the total number of lattice sites. The mean field energies 
$E_{\pm} = -\mu'({\bf k})\,\pm \,E_{q}({\bf k})$ 
are the same as those derived 
earlier in Ref. \cite{azzouz2003} when we let $b_0=UQ$; $Q$ 
being then the RAF order parameter satisfying 
Eq. (\ref{mean-field equation}). Here, $\epsilon_1({\bf k})=-2t(\cos k_x+\cos k_y)$
and $\mu'({\bf k})=-\mu-4t'\cos k_x\cos k_y+Un$. 
$n=\langle c_{i\sigma}\rangle$ is the electron's density, which satisfies
the following mean-field equation \cite{azzouz2003} 
\begin{eqnarray}
n&=&\frac{1}{2N}\sum_{\bf k}{n_F[E_+({\bf k})]+n_F[E_-({\bf k})]}.
\label{density}
\end{eqnarray}
Note that the decoupling the quartic interacting term of the Hubbard 
Hamiltonian using this density order parameter led to adding 
$Un$ in the expression of $\mu'({\bf k})$, \cite{azzouz2003}.
The fluctuations beyond the mean-field solution are considered
for the RAF order only for simplicity.
Also, the fluctuations considered here are in the longitudinal direction
of the RAF parameter, since we argue that these are
much more important than the transverse fluctuations, given that 
the phase of the local RAF parameter is time dependent, 
so already fluctuating at the 
mean-field level.

Upon Fourier transforming to ${\bf k}$ and frequency space, 
the mean-field action takes on the form

\begin{equation}
S_{0}=\sum_{\tilde k}\psi_{\tilde k}^* 
{\cal G}^{-1}\psi_{\tilde k}+ N\frac{\mid b\mid^{2}}{U},
\end{equation}
where ${\tilde k}\equiv({\bf k},\omega_n)$; 
${\bf k}$ being the wavevector and $\omega_n$
the fermionic Matsubara frequency. Here,
$ \psi_{\tilde k}^*=(c_{{\tilde k}\uparrow}^{A^*}\ 
c_{{\tilde k}\uparrow}^{B^*}\ c_{{\tilde k}\downarrow}^{A^*}\ 
c_{{\tilde k}\downarrow}^{B^*})
$ is a 4-component spinor, 
and the mean-field Green's function is \cite{azzouz2005}
\begin{equation} 
{\cal G}
({\bf k},i\omega_{n})=\frac{[i\omega_{n}+
\mu^{\prime}({\bf k})]I+\epsilon({\bf k}){\cal M}+b{\cal N}}
{[i\omega_{n}+\mu^{\prime}({\bf k})]^2-[\epsilon^2({\bf k})+b^2]},
\label{greensfunction}
\end{equation}
with
\begin{eqnarray}
\mathcal{\cal M} =
\left(\begin{array}{cc}
\tau_1 & 0 \\
0 & \tau_1
\end{array} \right),
\ \ \ \ 
\mathcal{\cal N} =
\left(\begin{array}{cc}
0&\tau_3 \\
\tau_3& 0
\end{array} \right),
\end{eqnarray}
where $\tau_1$ and $\tau_3$ are the first and third Pauli matrices.

In order to go beyond the mean-field solution, we consider 
the Gaussian fluctuations by writing
\begin{equation}
b_i = b_0 + \delta b({\bf {r_i}},\tau),
\end{equation}
with $\delta b({\bf {r_i}},\tau)$ a small deviation around the mean-field point.
Using the approach for calculating Gaussian contributions to 
the partition function described in Ref. \cite{negele} one finds
\begin{equation}
Z=Z_0\,\int \prod_i d (\delta b_i) 
\exp\bigg(-\frac{1}{2}\int_{0}^{\beta}\,
d\tau\sum_{i}\delta b_i \Gamma^{-1}\delta b_i^*\bigg), 
\end{equation}
where $Z_0$ is the mean-field partition function, and
$\Gamma$ the propagator of the Gaussian fluctuations, given in Fourier space by
\begin{equation}
\Gamma({\tilde q})=\frac{2U}{1-{U} \chi({\tilde q})/4}.
\end{equation}
The particle-hole type bubble $\chi$ reads as
\begin{eqnarray}
\chi({\tilde q}) &=& \sum_{\omega_n}\int\frac{d^{2}k}{(2\pi)^2}{\rm Tr} 
[{\cal G}({\tilde k})\,{\cal N} 
{\cal G}({\tilde k}+{\tilde q})\,{\cal N}] \cr
&=&\int\frac{d^{2}k}{(2\pi)^2}  \frac{d\epsilon 
d\epsilon^\prime}{(2\pi)^2}\frac{n_F(\epsilon)-
n_F(\epsilon^\prime)}{\epsilon - \epsilon^\prime + i\omega_m} \cr
&&{\rm Tr}[A({\bf k},\epsilon){\cal N} A({\bf k}
+{\bf q},\epsilon^\prime){\cal N}].
\end{eqnarray}
where ${\tilde q}\equiv({\bf q},i\omega_m)$; $\omega_m$ 
being the bosonic Matsubara frequency, and ${\rm Tr}$ 
designates the trace of a matrix. 
The spectral function $A({\bf k},\epsilon)$ is 
related to the Green's function by $
{\cal G}({\bf k},i\omega_n) =  \int \frac{d\epsilon}{2\pi} 
\frac{A({\bf k},\epsilon)}{i\omega_n - \epsilon}$.
The imaginary part of $\chi$, 
which is needed in the calculation of the imaginary part of the self-energy, is
\begin{eqnarray}
\chi''(q,\omega)&=& \int \frac{d^{2}k}{(2\pi)^2} \frac{d\epsilon}{4\pi}
[n_F(\epsilon)-n_F(\epsilon + \omega)]
{\rm Tr}[A({\bf k},\epsilon){\cal N} 
A({\bf k}+{\bf q},\epsilon+\omega) {\cal N}] \cr
&\approx&
\int\frac{d{\bf k}}{(2\pi)^2}\frac{\omega}{\pi}
\sum_{s=\pm}\sum_{s'=\pm}C_{ss'}({\bf k},{\bf q})
\frac{\eta}{[\mu'({\bf k})+sE_q({\bf k})]^2+\eta^2} \cr
&&\times \frac{\eta}{[\omega+\mu'({\bf k}-{\bf q})+s'E_q({\bf k}-{\bf q})]^2+\eta^2}
\label{chipp}
\end{eqnarray}
where 
$$
C_{ss'}({\bf k},{\bf q})=1+ss'\frac{U^2Q^2-\epsilon({\bf k})\epsilon({\bf k}-{\bf q})}
{E_q({\bf k})E_q({\bf k}-{\bf q})}.
$$
In (\ref{chipp}), we used $n_F(\epsilon)-n_F(\epsilon + \omega)\approx
\omega\delta(\epsilon)$ in the low frequency and low temperature regime.
In order to derive an expression for $\chi''$ consistent 
with a memory function-like approximation \cite{ope2000,gotze1972},
the $\omega$ independent Lorentzian in (\ref{chipp}) is replaced
by a delta function in the limit $\eta\to0$:
$
\frac{\eta}{[\mu'({\bf k})+sE_q({\bf k})]^2+\eta^2}
\approx\pi\delta(\mu'({\bf k})+sE_q({\bf k})).
$
This constrains the integration over ${\bf k}$ to be performed
over the FS, which satisfies 
$\mu'({\bf k})+sE_q({\bf k})=0$ with $s=\pm$; 
Ref. \cite{azzouz2010p}.
With this, $\chi''$ assumes the simpler form
\begin{eqnarray}
&&\chi''(q,\omega)=
\int_{FS}\frac{d{\bf k}}{(2\pi)^2}\sum_{s,s'} \cr
&&\frac{C_{ss'}({\bf k},{\bf q})\omega\eta}{[\omega+
\mu'({\bf k}-{\bf q})+s'E_q({\bf k}-{\bf q})]^2+\eta^2}.
\label{chipp2}
\end{eqnarray}
The integral in (\ref{chipp2})
runs over points belonging in the FS, only. 
In the high temperature limit or in the overdoped regime,
the FS in RAFT consists of large contours around 
$(0,0)$ and $(\pi,\pi)$ \cite{azzouz2013,azzouz2010}. Also the FS surface in this case 
is characterized by significant nesting for momenta transfers
slightly different than $(\pi,\pi)$ \cite{azzouz2010,azzouz2013}.
This nesting property is significantly reduced in the
underdoped regime below $T^*$, because the FS reconstructs into
small hole pockets around the points $(\pm\pi/2,\pm\pi/2)$ \cite{azzouz2010,azzouz2013}. 
The presence of the PG in this case also reduces the density of states
for wavevectors on the FS. These facts will 
cause $\chi''$ to be greater in the overdoped regime
and for temperatures greater than $T^*$ in the underdoped regime in general.
This behavior of $\chi''$ will affect the doping and temperature dependence
of self-energy as explained next.

\section{Derivation of self-energy in the presence of the PG}
\label{sec3}

\begin{figure}
		\includegraphics[height=5.5cm, angle=-90]{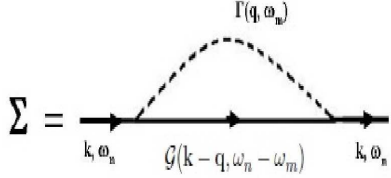}
\caption{The leading order self-energy diagram is drawn. 
The dashed line is the propagator 
	${\Gamma}$ of 
	the Gaussian fluctuations beyond the mean-field solution. The continuous line
	is the propagator ${\cal G}$ of the quasi-particles in the mean-field solution.}
  \label{Fig1}
\end{figure}

Using the Feynman diagram for self-energy depicted in Fig. \ref{Fig1}, we write
\begin{eqnarray}
\label{self energy}
\Sigma({\tilde k}) = 2T\int \, \frac{d^2 q}{(2\pi)^2} \,\sum_{\omega_m} {\cal G}
({\tilde k}-{\tilde q})\,\Gamma({\tilde q}).
\end{eqnarray}
In order to carry on the calculations for self-energy 
we expend $\Gamma({\tilde q})$ to second order in $U$ 
in the limit of $U < W$, with $W$ being the bare 
bandwidth energy ($W=8t$ if $t'=0$). Keeping only the lowest-order 
term contributing to the imaginary part of 
self-energy one gets
\begin{eqnarray}
\Sigma({\tilde k})&\approx&  U^2\int \frac{d^2 q}{(2\pi)^2}  
\frac{d^{2}k^\prime}{(2\pi)^2}\frac{d\epsilon}{2\pi} 
\frac{d\epsilon^\prime}{2\pi} \frac{d\epsilon^{''}}{2\pi} \Delta n_F 
A({\bf k}-{\bf q},\epsilon^{''})\cr
 &&\times{\rm Tr}[A({\bf k}^\prime,\epsilon){\cal N} 
A({\bf k}^\prime+{\bf q},\epsilon^\prime) {\cal N}]
\frac{n_F (- \epsilon^{''}) + n_B (\epsilon^\prime - \epsilon)}
{i\omega_n - \epsilon^{''} - \epsilon^\prime + \epsilon}, \nonumber\\
\end{eqnarray}
where $\Delta n_F = n_F(\epsilon) - n_F(\epsilon^\prime)$, and $n_B$
is the Einstein-Bose factor.
Taking the analytical limit 
$i\omega_n \longrightarrow \omega + i0^+$ gives the following 
expression for the imaginary 
part of self-energy:
\begin{eqnarray}
\label{self energy2}
\Sigma''({\bf k,\omega}) 
&=&  U^2\int \frac{d^2 q}{(2\pi)^2} \frac{d^{2}k^\prime}{(2\pi)^2} 
\frac{d\epsilon d\epsilon^\prime}{8\pi^2} 
[{n_F (\epsilon'-\epsilon-\omega) + n_B (\epsilon^\prime - \epsilon)}]\cr
&&\Delta n_F 
A({\bf k}-{\bf q},\epsilon-\epsilon'+\omega)
{\rm Tr}[A({\bf k}^\prime,\epsilon){\cal N} 
A({\bf k}^\prime+{\bf q},\epsilon^\prime) {\cal N}].
\end{eqnarray}
In the limit of the tight-binding electrons 
where the PG is absent, so with
$A({\bf k}-{\bf q},\epsilon-\epsilon'+\omega)\approx 2\pi
\delta(\omega+\epsilon-\epsilon'-\epsilon_{{\bf k}-{\bf q}})$,
Eq. (\ref{self energy2}) is shown to reduce to 
the same expression as in the second-order 
Born approximation used in NAFL \cite{stojkovic1997}, namely:
\begin{eqnarray}
\Sigma''({\bf k},\omega)\approx g^2\int\frac{d^2k'}{4\pi^2}
\chi''({\bf k}-{\bf k}',\omega-\epsilon_{{\bf k}'})
[{n_F (\epsilon_{{\bf k}'}) 
+ n_B (\epsilon_{{\bf k}'}-\omega})],
\label{born approximation}
\end{eqnarray}
where $g=U$ in the present intermediate coupling regime.

It is possible to derive
a Born approximation that incorporates the PG effect 
using the mean-field result for the spectral function. The latter
is obtained from the Green's function (\ref{greensfunction}) through 
$A({\bf k},\omega)=-2{\rm Im}G({\bf k},\omega)$; Ref. \cite{azzouz2005}:
\begin{eqnarray}
A({\bf k},\omega)=\sum_{s=\pm}
\frac{\eta}{[\omega+\mu'({\bf k})+sE_q({\bf k})]^2+\eta^2}a_s({\bf k})
\label{MF spectral function}
\end{eqnarray}
with 
$a_s({\bf k})=
I-s\big[\frac{\epsilon({\bf k})}
{E_q({\bf k})}{\cal M} + 
\frac{UQ}{E_q({\bf k})}{\cal N}\big]$. $I$ is the $4\times4$ identity 
matrix.
Taking $\eta\to0^+$ gives
\begin{eqnarray}
A({\bf k},\omega)&=&\pi\sum_{s=\pm}
\delta\bigg(\omega+\mu'({\bf k})+sE_q({\bf k})\bigg)a_s({\bf k}).
\label{MF spectral function2}
\end{eqnarray}
The Dirac delta function in (\ref{MF spectral function2}) allows
the integration over $\epsilon'$ in (\ref{self energy2}) to be readily 
performed.  This results in a much simpler expression for
the imaginary part of self-energy: 
%
%
%
%
%
%
%
%
%
$
$
\begin{eqnarray}
\label{self energy3}
\Sigma''({\bf k,\omega}) 
&=&  \sum_{s=\pm}
U^2\int \frac{d^2 {k}'}{(2\pi)^2}  
a_s({\bf k}') 
\chi''\big({\bf k}-{\bf k}', \omega - E_s({\bf k}')\big) \cr
&&[n_F\big(E_s({\bf k}')\big) 
+ n_B\big( E_s({\bf k}')- \omega \big)].
\end{eqnarray}
There are two noticeable effects for the PG on self-energy. 
First, the tight-binding energies 
in the Fermi and Bose factors as well as in $\chi''$ are replaced by 
RAFT's eigenenergies $E_\pm({\bf k})$. Second, the self-energy becomes a matrix
when the PG is present, with the off-diagonal elements caused by the 
terms proportional to the matrices ${\cal M}$ and ${\cal N}$ 
in the RAFT's spectral function (\ref{MF spectral function2}).

As it is extremely difficult to calculate $\Sigma''$ analytically due to 
the double integral over ${\bf k}$ and to the dependence on $\chi''$, 
which is itself difficult to calculate analytically,
we performed this calculation numerically. Note that Stojkovic 
and Pines \cite{stojkovic1997} calculated analytically $\Sigma''$
in the absence of the PG using the gapless second-order 
Born approximation in
Eq. (\ref{born approximation}). They found that 
the ${\bf k}$-averaged scattering rate takes on the MFL
form. In our case, we also get this MFL behavior in addition to other effects
due to the PG, which were not included in Stojkovic 
and Pines' work.

\section{Results}
\label{sec4}

\begin{figure}
    \includegraphics[height=5.5cm]{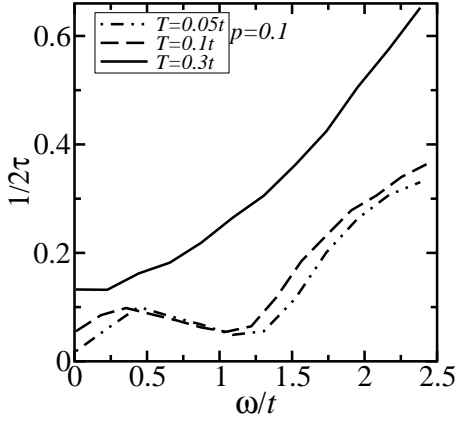}
  \caption{$1/2\tau=-\Sigma''$ is plotted versus $\omega$ for the wavevector
	$(k_x,k_y)=(k_F,k_F)\in$ FS with $k_F=0.4\pi$. 
	Temperature and doping are indicated on the figure.
	The Hamiltonian parameters used in the present work 
	are $U=3t$ and $t'=-0.16t$.
	All numerical calculations were performed on a $100\times100$
	Brillouin zone, and a mesh of $100\times100$ points for the momentum transfer. 
	Here, ${\bf k}=(0.4\pi,0.4\pi)$ along the diagonal.
$1/\tau$ is in units of $t$.}
  \label{Fig2}
\end{figure}

\begin{figure}
    \includegraphics[height=5.5cm]{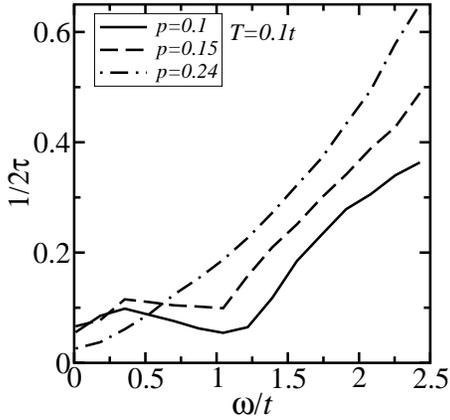}
  \caption{$1/2\tau=-\Sigma''$ is drawn versus frequency 
	$\omega$ for three values of doping at temperature $T=0.1t$.
	Here, ${\bf k}=(k_F,k_F)\in$ FS along the diagonal, with $k_F=0.4\pi$. 
Note that the Fermi wavevector does not change significantly 
with doping along the diagonal.}
  \label{Fig3}
\end{figure}

Figure \ref{Fig2} displays ${1}/{2\tau}=-\Sigma''$ versus the frequency 
$\omega/t$ 
for the wavevector ${\bf k}_{F1}=(0.4\pi,0.4\pi)$ on the FS and for
three different temperatures and a fixed doping $p=0.1$. 
The Hamiltonian parameters are $U=3t$, and $t'=-0.16t$.
Here, for the sake of simplicity,
we focus only on the diagonal elements of the self-energy, which
are all equal.
At the highest temperature $T=0.3t$, 
${1}/{2\tau}$ shows a mixture of FL and MFL behaviors.
$1/2\tau$
can be fitted using linear and quadratic
terms in $\omega/t$. 
The linear frequency dependence (MFL) is a consequence of 
the nesting property of the FS. There is no PG 
at this temperature, and the FS consists 
of large contours around $(0,0)$ and $(\pi,\pi)$. 
For lower temperatures ($T=0.1t$ and 
$T=0.05t$), the PG is present, and the FS reconstructs into 
pockets around $(\pm\pi/2,\pm\pi/2)$ \cite{azzouz2013,azzouz2010p}. 
The nesting surface shrinks and the number of quasiparticle states 
available in the system becomes smaller. 
The first consequence of the nesting decrease 
is to reduce ${1}/{2\tau}$ by roughly a factor $2$ at high 
frequencies. At low frequencies,  the electron-electron scattering
processes are reduced because of the depletion of the density of 
states at the Fermi energy
due to the PG, and the quasiparticle lifetime $\tau$ increases. 
This effect is more pronounced at lower temperature.

Figure \ref{Fig3} displays ${1}/{2\tau}$ versus $\omega/t$ 
for different doping values at the temperature $T=0.1t$ 
and for the wavevector ${\bf k}_{F1}=(0.4\pi,0.4\pi)$ on the FS.
In the underdoped regime  with $p=0.1$ and $0.15$, ${1}/{2\tau}$ 
shows a linear behavior at high values of $\omega/t$  and a 
downward deviation at lower frequencies, which occurs at a value of 
$\omega/t$ that increases when doping decreases. This is a 
signature of the PG, which is bigger at lower doping.  
For  $p=0.24$ in the overdoped regime, 
${1}/{2\tau}$ can be fitted by linear and quadratic terms
in $\omega/t$. Again, this is  
a mixture of FL and MFL behaviors, but 
${1}/{2\tau}$ goes to a lower value when $\omega/t$ tends 
to zero indicating a greater FL tendency.

\begin{figure}
    \includegraphics[height=5.5cm]{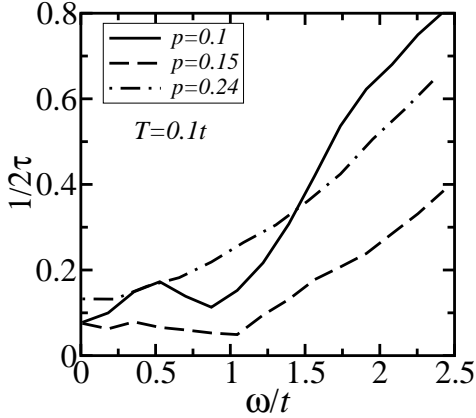}
  \caption{$1/2\tau=-\Sigma''$ is drawn versus frequency 
	$\omega$ for three values of doping at temperature $T=0.1t$ and for wavevectors on the 
	FS.
	The wavevectors are 
${\bf k}=(0.47\pi,0.41\pi)$ for $p=0.1$,
${\bf k}=(0.74\pi,0.71\pi)$ for $p=0.15$,
and ${\bf k}=(0.45\pi,0.4\pi)$ for $p=0.24$. 
Here, the Fermi wavevector changes 
significantly with doping away from the diagonal.}
  \label{Fig4}
\end{figure}

In figure \ref{Fig4}, $1/2\tau$ is drawn 
versus $\omega/t$
for three doping levels and for wavevectors
on the FS away from the diagonal of the Brillouin zone. 
Due to the FS reconstruction with doping, 
these wavevectors are different for different
dopings. First of all, it is clear from this figure and figure
\ref{Fig3} that $1/2\tau$ shows 
a strong ${\bf k}$ dependence. For example,
for $p=0.1$ in the underdoped regime, 
there is a factor of $2$ roughly between 
$1/2\tau$ for the FS point away from the diagonal (Fig. \ref{Fig4})
and the FS point on the diagonal (Fig. \ref{Fig3}).
For this doping ($p=0.1$),
the MFL linear behavior at high frequencies 
is followed by a deviation from linearity 
as the frequency decreases, then by a hump due to the PG 
at even lower frequencies.
For $p=0.15$ near the optimal point,
$1/2\tau$ decreases linearly with frequency, then 
saturates for $\omega\leq t$ and presents 
a hump due to the PG at even lower frequency.
These trends are the same as those encountered for 
the FS point on the diagonal of the Brillouin zone
in Fig. \ref{Fig3}.
For the doping $p=0.24$ above the optimal point 
(where the PG is zero),
$1/2\tau$ can be fitted using linear and quadratic 
terms; it thus shows a mixture of FL and MFL behavior
like for the FS point on the diagonal.

Regarding the comparison with experiment,
our results for self-energy
are qualitatively consistent with the experimental data 
of the optical conductivity, which was analyzed using the extended
Drude model \cite{timusk1999}. The downward deviation from linearity
in $1/2\tau$ versus frequency, 
observed experimentally, is accounted for 
in the present theory. 
Taking $t=0.1$ eV, which is the value considered in RAFT's past works,
we find that the values of $1/\tau$ are in the same range
as the experimental ones \cite{timusk1999}.
Note that $1/2\tau$ is in units of $t$ in figures 
\ref{Fig2}, \ref{Fig3}, and \ref{Fig4}. 
There are however 
discrepancies between the calculated relaxation rate and the 
experimental one.
These differences can be
attributed mainly to the following reasons:
the extended Drude analysis used the Drude conductivity with
a mass enhancement factor and a frequency and temperature dependent
relaxation rate. This analysis does not take into 
account the PG explicitly, contrary to the 
present microscopically calculated 
self-energy, which does include the PG.
Also, in RAFT, the establishment of the PG below $T^*$
in the underdoped regime causes the reconstruction of the FS. 
This important property 
is not unfortunately
taken into account in the extended Drude analysis.
Note that our present approach does not consider the
contributions of the real part of self-energy
in this comparison with
the extended Drude analysis.

\section{Conclusions}
\label{sec5}

In order to justify the usage of a MFL-like self-energy
in past works based on RAFT,
we calculated the self-energy 
corrections beyond the mean-field solution 
of this theory, and found that
the doping, temperature and frequency dependences of this self-energy 
agree qualitatively well with the results of the nearly antiferromagnetic Fermi 
liquid as far as this MFL behavior is concerned.
Also, the trends of the imaginary part of this self-energy
capture well the main features of the relaxation rate derived 
in the extended Drude analysis of the optical conductivity.
Note that, contrary to that analysis,
the expression of self energy in the present work depends
explicitly on the pseudogap. The main consequences of the 
latter are a deviation from linearity and a hump 
in the low frequency regime in the frequency dependence
of the relaxation rate $1/\tau$. According to the 
present results,
the changes in the frequency dependence of self-energy
as doping goes from the overdoped regime to the underdoped regime
are due to the reconstruction of the Fermi surface near optimal doping.

\section*{Acknowledgement}
M.A. would like to thank the Laboratoire de Physique de la Mati\`ere Condens\'ee 
and the department of physics at Al Manar University 
in Tunisia for their hospitality during his visits that led to 
the completion of this project.


\begin{thebibliography}{30}

\bibitem{timusk1999} T. Timusk, B. Statt, Rep. Prog. 
Phys. {\bf 62}, 61 (1999).

\bibitem{bednorz1986} J.G. Bednorz and K.A. Mueller, Z. Phys. 
B- Condensed Matter {\bf 64}, 189 (1986).

\bibitem{he2011} Rui-Hua He, M. Hashimoto, H. Karapetyan, J.D. Koralek, 
J.P. Hinton, J.P. Testaud, V. Nathan, Y. Yoshida, Hong Yao, 
K. Tanaka, W. Meevasana, R.G. Moore, D. H. Lu,
S.-K. Mo, M. Ishikado, H. Eisaki, Z. Hussain, T.P. Devereaux,
S.A. Kivelson, J. Orenstein, A. Kapitulnik, Z.-X. Shen,
Science {\bf 331}, 1579 (2011).

\bibitem{chakravarty2001} S. Chakravarty, R. B. Laughlin, D. K. Morr, and C. Nayak, Phys.
Rev. B {\bf 63}, 094503 (2001). This paper proposed the so-called DDW order to 
be in competition with d-wave superconductivity.

\bibitem{azzouz2003}M. Azzouz, Phys. Rev. B {\bf 67}, 134510 (2003).

\bibitem{azzouz2003p}M. Azzouz, Phys. Rev. B {\bf 68}, 174523  (2003).

\bibitem{azzouz2004}M. Azzouz, Phys. Rev. B {\bf 70}, 052501  (2004).

\bibitem{azzouz2012p} M. Azzouz, Physica C {\bf 480}, 34  (2012).

\bibitem{azzouz2013}M. Azzouz, Spectrum {\bf 5}, 215 (2013).

\bibitem{azzouz2013p} M. Azzouz, in preparation (2013).

\bibitem{azzouz2005}H. Saadaoui and M. Azzouz, 
Phys. Rev. B {\bf 72}, 184518 (2005).

\bibitem{azzouz2012} E.H. Bhuiyan, G. Presenza-Pitman, M. Azzouz,
Physica C {\bf 473}, 61 (2012).

\bibitem{azzouz2010} M. Azzouz, K.C. Hewitt, H. Saadaoui, 
Phys. Rev. B
{\bf 81}, 174502 (2010).

\bibitem{azzouz2010p} M. Azzouz, B.W. Ramakko, G. Presenza-Pitman,
J. Phys.: Condens. Matter {\bf 22}, 345605 (2010).



\bibitem{varma1989} C. M. Varma, P. B. Littlewood, and S. Schmitt-Rink,
E. Abrahams and A. E. Ruckenstein, Phys. Rev. Lett. {\bf 63}, 1996 (1989)





\bibitem{stojkovic1997} B.P. Stojkovic and D. Pines, Phys. 
Rev. B {\bf 56}, 11931 (1997).


\bibitem{negele} J.W. Negele and H. Orland, Quantum Many-Particle Systems, 
Addison-Wesley Publishing Company (1988). 


\bibitem{ope2000}
M. Opel, R. Nemetschek, C. Hoffman, R. Philipp, P.F. M$\ddot{\mbox{u}}$ller,
R. Hackl, I. T$\ddot{\mbox{u}}$tto, A. Erb, B. Revaz, E. Walker, H. Berger,
and L. Forr\'o, Phys. Rev. B {\bf 61},  9752  (2000).

\bibitem{gotze1972} W. G$\ddot{\mbox{o}}$tze and P. W$\ddot{\mbox{o}}$lfe,
Phys. Rev. B {\bf 6}, 1226 (1972).







\end{thebibliography}
\end{document}